\documentclass[letterpaper, conference]{IEEEtran}

\usepackage{amsmath, amsfonts, amsthm}
\usepackage{authblk}
\usepackage{bm}
\usepackage{booktabs}
\usepackage[ruled,vlined,linesnumbered]{algorithm2e}
\usepackage{enumitem}
\usepackage{subfig}
\usepackage{tikz-network}
\usepackage{url}
\usepackage{graphicx}

\usepackage[utf8]{inputenc}
\usepackage[english]{babel}

\title{\change{Smart Advertisement for Maximal Clicks in Online Social Networks Without User Data}}
\author[1]{Nathaniel Hudson}
\author[1]{Hana Khamfroush}
\author[1]{Brent Harrison}
\author[2]{Adam Craig}
\affil[1]{Department of Computer Science, University of Kentucky}
\affil[2]{Department of Marketing \& Supply Chain, University of Kentucky}

\newcommand{\needref}[1]{\textcolor{red}{\textbf{[ref]}}}

\newcommand{\change}[1]{#1} 

\DeclareMathOperator*{\argmax}{argmax}

\begin{document}
\maketitle

\begin{abstract}
    Click-through rate (CTR) prediction of advertisements on online social network platforms to optimize advertising is of much interest. Prior works build machine learning models that take a user-centric approach in terms of training -- using predominantly user data to classify whether a user will click on an advertisement or not. While this approach has proven effective, it is inaccessible to most entities and relies heavily on user data. To accommodate for this, we first consider a large set of advertisement data on Facebook and use natural language processing (NLP) to extract key concepts that we call conceptual nodes. To predict the value of CTR for a combination of conceptual nodes, we use the advertisement data to train four machine learning (ML) models. We then cast the problem of finding the optimal combination of conceptual nodes as an optimization problem. Given a certain budget $k$, we are interested in finding the optimal combination of conceptual nodes that maximize the CTR. A discussion of the hardness and possible NP-hardness of the optimization problem is provided. Then, we propose a greedy algorithm and a genetic algorithm to find near-optimal combinations of conceptual nodes in polynomial time, with the genetic algorithm nearly matching the optimal solution. We observe that Decision Tree Regressor and Random Forest Regressor exhibit the highest Pearson correlation coefficients w.r.t. click predictions and real click values. Additionally, we find that the conceptual nodes of ``politics", ``celebrity", and ``organization" are notably more influential than other considered conceptual nodes.
\end{abstract}



\section{Introduction}
\change{Smart cities are becoming an increasingly important paradigm as we design future technological systems. As the interdependency between technology and resources managed by cities become more intertwined, the need to manage them intelligently will be exacerbated~\cite{kakarontzas2014conceptual, yin2015literature, arasteh2016iot, pellicer2013global}. The analytics of human activity will also become an important facet of smart cities. Smart decisions, empowered by technology via sensor networks and real-time analytics, made in response to human social activity is on the horizon. For instance, if there is a sizable political protest in a given area, a smart city could use real-time analytics to recognize that and then make smart traffic routing decisions so that the protest causes minimal traffic congestion~\cite{doran2013human, anastasi2013urban}. Smart advertisement to more effectively distribute information based on social trends sensed from smart city settings is the point of interest for this paper. For this work, we consider social media data.}

Social media platforms allow for accessible, immediate, and wide-spanning interactions with people across the globe --- allowing for social interaction not previously possible. With the wide adoption of these platforms, much effort is made, in part, by businesses and corporations to market products on these platforms to reach potential customers~\cite{SrivastavaAvikalp2017SMAO, ad1, ad2}.
Increasingly, political organizations and outside nations may also seek to influence political attitudes and voting behavior,
using these platforms~\cite{LaneDanielS2017FODt}.
In order to increase their influence across a platform, stakeholders will target certain interests, demographics, topics, etc. to promote their product or cause~\cite{ad3, ad4}. 
With this vested need to try to more effectively market information and products to users of these platforms, there is an interest in understanding how and why people interact with certain information on these social media platforms~\cite{ctr-seminal1, ctr-seminal2}. A better understanding of how to design advertisements to attract more clicks from users has many practical applications. For instance, this can be used to more effectively propagate vital information throughout a population or to better sell a product. It could also give us some insight as to how misinformation spreads on social media platform so that further research can be committed to preventing such phenomena. 

Click-through rate (CTR) prediction is a domain of research invested in developing systems that can predict how users will respond, in the measure of clicks, to given content (though usually advertisements). Contextual advertising is where advertisements are placed in contexts specific to an individual user, so decisions as to which advertisement to display in CTR prediction is often based on user data~\cite{ctr-facebook, ctr-twitter, ctr-alibaba}. These works rely heavily on having direct access to rich user data to train their learning models and improve their accuracy over time.
Some of the state-of-the-art works modeling approaches to these problems are described in detail in Section~\ref{sec:related-works}. However, for our work, we are more interested in understanding how features related to the advertisements themselves can be used to perform CTR prediction, and which features attract more CTR.
From there, this work studies selecting the optimal combination of targeted interests as an optimization problem and proposes algorithms to maximize expected clicks using an approach to CTR prediction that does not touch sensitive user data and maintains user privacy.
Additionally, there is interest as to which topics, themes, sentiments, etc. drive more people to click advertisements. 

\change{Our approach to performing CTR prediction does not include user data {(e.g., names, demographics, age, location, user connectivity)} that could be sensitive. 
{Amidst growing concerns of privacy and distrust of large entities handling user data, 
there is a need for new approaches to performing such tasks such that they no longer rely on user data.}
The need for smart advertisement is not going to dampen to satisfy privacy concerns. So, for this work, we propose the notion of {\it conceptual nodes} to thematically group targeted interests (e.g., ``2nd amendment" and ``guns") that are used to specify the users an advertiser wishes to reach into holistic groups (e.g., ``politics"). 
{These conceptual nodes will be used to perform CTR prediction for advertisements based on their combinations of conceptual nodes. To clarify, conceptual nodes will be based entirely on the content of the designed advertisement.}
We then use machine learning models to learn which conceptual nodes are most effective at attracting clicks from users. Once trained, these models will be used to perform smart advertisement by {\it intelligently selecting some optimal set of conceptual nodes, under a budgeting constraint $k$ for how many conceptual nodes an advertisement can have, to maximize the expected number of clicks for an advertisement}. We refer to this problem as the 
{\it Optimal Conceptual Node Combination} (OCNC) problem. Our results from this work show that this approach to CTR prediction can demonstrate reasonable efficacy despite the lack of user data to inform the machine learning model, as found in state-of-the-art approaches to CTR prediction.}


To support this work, we use data provided by the U.S. Congress. These data contain controversial, misleading, and/or hyperbolic advertisements distributed on Facebook as part of Russia's (mis)information campaign leading up to and after the U.S. 2016 presidential election. These data containing advertisements purchased by the Internet Research Agency~(IRA) --- a notorious Russian ``troll" farm --- were made freely and publicly available by the U.S. Congress~\cite{Data}.  
To the best of our knowledge, this is the only investigative study of CTR prediction using this data-set.
Using these data, we perform a novel approach to CTR prediction. What separates our method for CTR prediction from the state-of-the-art CTR prediction models (in contextual advertising) is the lack of access to user data. For this work, we are not interested in asking whether a user will click on an advertisement or not; instead we are interested in understanding which features of advertisements lead to high CTR. The contributions of this paper can be summarized in the following points:
\begin{itemize}
	\item Model the problem of CTR maximization as an optimization problem (OCNC problem) and discuss the NP-hardness of OCNC.
	\item Introduce the notion of {\it conceptual nodes} to encapsulate themes of targeted interests to perform CTR prediction whilst preserving user privacy by not using user data. 
	\item Propose efficient genetic and greedy algorithms to optimally select conceptual node combinations to maximize expected clicks of advertisements. 
	\item Social insights on which conceptual nodes were the most effective at attracting users to click advertisements on Facebook. These insights provide some \change{idea} on potentially effective advertisement strategies.
\end{itemize}

\section{Related Works} \label{sec:related-works}
Much work has been devoted to predicting click-through rates (CTR) of online advertisements. In this domain of study, there are typically two different flavors of online advertisements considered: {\it sponsored search} and {\it contextual advertising}. The former deals with static webpages and placing advertisements w.r.t. to a search query provided by a user~\cite{ctr-seminal1, ctr-seminal2}. The latter deals with more user-centric approaches that select advertisements that are more appropriate for a given user based on historic data, categorized interests, etc. When considering CTR prediction for advertisements in social media platforms, contextual advertising is more appropriate.

For contextual advertising on social media platforms, much work has been done to accurately predict CTR for advertisements. He et al. in~\cite{ctr-facebook} considered interests and demographics of users on Facebook for their classifier models that are built on top of decision tree and logistic regressor models. Their model is then incorporated in a recurrent architecture to continually improve accuracy over time as the model continues to get feedback from users. The authors in~\cite{ctr-twitter} introduced a learning-to-rank model to predict CTR of advertisements placed in a unique online Twitter stream, composed of Tweets shared by a user's followees. Predictions made by this model are based on user-specific input features. Researchers at Alibaba in~\cite{ctr-alibaba} introduced a novel Deep Interest Network model that uses historical user data that adaptively learns user interests over time to improve CTR prediction.
A central commonality of these works is that they all are {\it user-centric} --- where their model is most interested in classifying whether an individual user will click a given advertisement or not. The models these works describe all rely on user data (interests, demographics, historical data, etc.) to predict whether a given user will click an advertisement before deciding whether to display the advertisement for the user under consideration. 

While these works are very interesting and make great steps forward for CTR prediction for contextual advertising, for our work we are interested in conducting an {\it content-aware} approach. Rather than investigate how to best predict whether a user will click an advertisement based on user features, we are interested in predicting how many clicks an advertisement will receive based on input features specific for that advertisement's content, targeted interests, etc., while ignoring user data entirely. This work is motivated by the increasing sensitivity surrounding how user data is used by OSN platforms to support many necessary for these services. We are additionally motivated to gain high-level insights to better understand what interests lead to higher CTR across advertisements and, more generally, across information shared on online social media platforms.

\begin{figure*}
    \centering
    \includegraphics[width=0.85\textwidth]{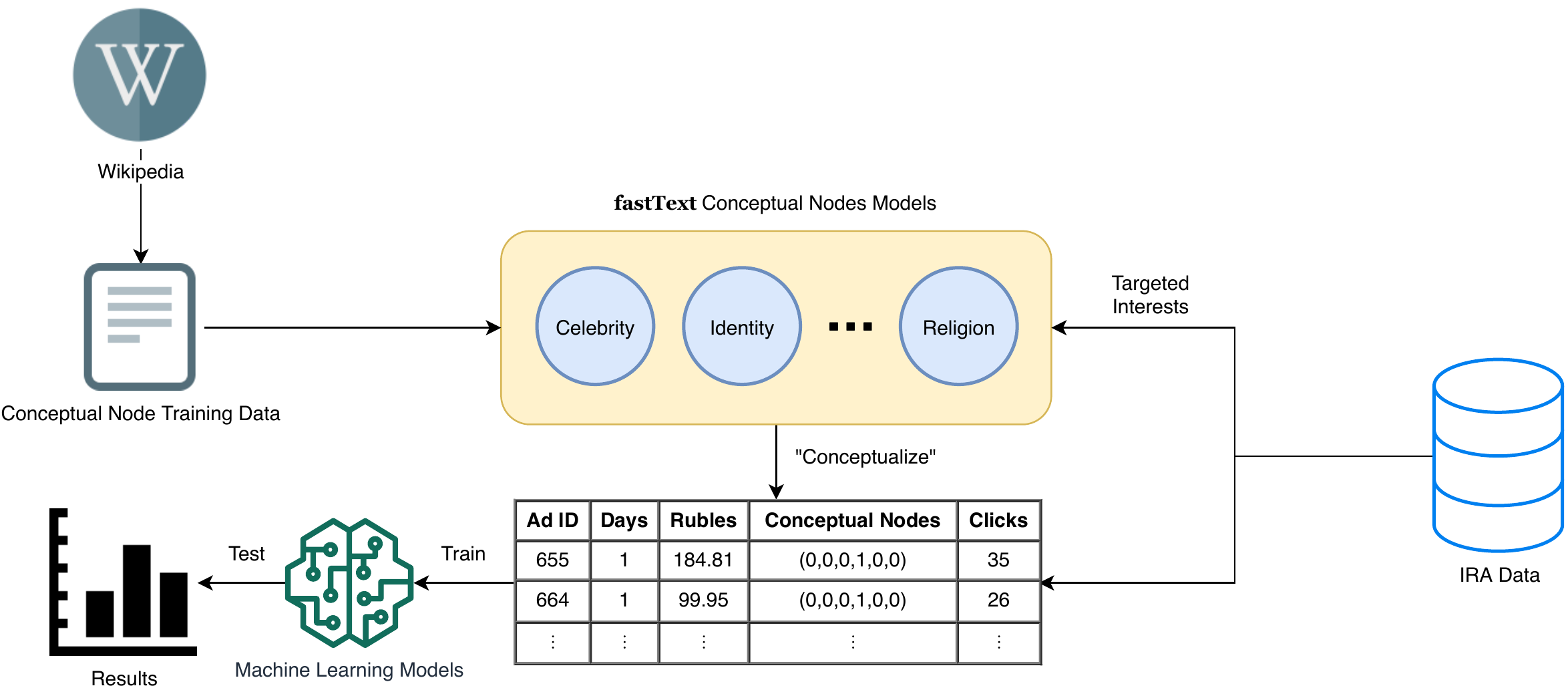}
    \caption{\change{Overview of the methodology used to perform (and validate) our approach to smart advertisement.}}
    \label{fig:process}
\end{figure*}

\section{Data Description}\label{sec:data-description}
For this work, we study and analyze data provided by the U.S. Congress --- which, from now on, we will refer to as the IRA data. Here, we provide some important measures of the data and describe our interest in it. The IRA data contains 3,519 PDF documents, with each PDF document containing information pertaining to a single advertisement. More than 11.4 million Americans were exposed to the advertisements featured in the IRA data~\cite{russian-ads} that we will be using for the bedrock of our analysis.

There are dozens of different features that are listed across the advertisements in the IRA data. {The set of features provided for each advertisement in the IRA data is not uniform. To compensate for that, we select a subset of features that are commonly found on most, if not all, advertisements. We then perform optical character recognition~(OCR) across the entire IRA data to create a single tabular data-set. This pre-processing made the data much more accessible than the original collection of PDF documents.} The features we consider for this work are: {\it clicks}, {\it spend}, {\it start\_date}, {\it end\_date}, and {\it targeted interests}. {While most of these listed features are included for every advertisement, {\it start\_date} and {\it end\_date} are only included in the majority.} Below is a brief {description} of each feature:
\begin{itemize}
	\item {\it clicks}: the integer representing the number of clicks an advertisement received.
	\item {\it spend}: the amount of monetary funding an advertisement received to increase its promotion on Facebook, currency is Russian rubles.
	\item {\it start\_date}: the date in which an advertisement began to be available online.
	\item {\it end\_date}: the date in which an advertisement would be last available online.
	\item {\it targeted interests}: a curated list {interests used to target certain people}, an important feature which lies as the foundation of our conceptual nodes, described in Section~\ref{sec:conceptual-nodes}.
\end{itemize}

\section{Conceptual Nodes} \label{sec:conceptual-nodes}
As mentioned in Section~\ref{sec:data-description}, each advertisement in the IRA data provides some important features to consider. One of the most interesting features to consider is {\it targeted interests}. This feature provides a list of curated interests that allow an advertisement {to} focus on people with interests in this list. For analysis, we wish to abstract the wide variety of targeted interests the entirety of this data-set includes. To do this, we introduce the notion of {\it conceptual nodes}. We consider conceptual nodes to essentially serve as categorical markers for these interests. 
{Using individual targeted interests to train machine learning models to predict CTR would be too granular. So, we use conceptual nodes to provide some level of abstraction and, essentially, clump together targeted interests into groups --- irrespective of how (in)frequently a targeted interest appears in the IRA data.}
Under this construct, ``LGBTQ+" and ``African American" can both belong to the same conceptual node because of their possible thematic similarities. For this work, we consider a fixed set of conceptual nodes. The {set of conceptual nodes considered for this work $\mathcal{N}$} are as follows: {\it celebrity}, {\it identity}, {\it news}, {\it organization}, {\it politics}, and {\it religion}. We chose these conceptual nodes to be general enough to accommodate the diversity of the targeted interests upon reading through a large number of advertisements in the IRA data. Further, these conceptual nodes were decided upon experimenting with different sets of conceptual nodes and seeing that these generally performed well at encapsulating the targeted interests --- of course, this could always be reconsidered for future works.

{To map targeted interests to} conceptual nodes, we employ natural language processing (NLP) models to learn word embeddings to approximately {map targeted} interests to appropriate conceptual nodes. The NLP model we employ for this work is Facebook's {\it fastText} model~\cite{fastText1, fastText2}. This model is an extension of the well-studied Word2Vec model~\cite{word2vec2}. 
Both models are unsupervised learning models, meaning that they can learn and recognize semantic features of words without the extra step of someone manually providing labeled data. An advantage {that} fastText has over Word2Vec is that it decomposes words it does not immediately recognize into smaller $n$-grams such that it can then approximate semantic values. For instance, if a fastText model's vocabulary does not include the word ``colors", but it does contain the word ``color", the fastText model will then break up the word into the following $n$-grams to calculate an approximate semantic value: ``color" and ``s".
To generate a large set of training {corpus}, we employ a recursive scraper we call $\mu${\it -Scraper}. This approach to scraping text data is elaborated in detail in Section~\ref{sec:scraper}.

Each considered conceptual node will have a fastText {model trained} with its own training {corpus}. Upon having a {trained} fastText model for each conceptual node, a targeted interest $i$ will be passed into each conceptual node's model to generate a value of semantic similarity.  {A targeted interest is then mapped to the conceptual node whose fastText model produces the highest semantic similarity value.}
For a visual overview of the method to incorporate conceptual nodes for click prediction, refer to Figure~\ref{fig:process}.
    
\subsection{{Textual Data Collection}} \label{sec:scraper}
In order to collect a large amount of pertinent natural language data to train a fastText model, we use a recursive {scraper} that strips text data from Wikipedia articles using their open-source Python API\footnote{\url{https://pypi.org/project/wikipedia/}}. The procedure for the scraping algorithm takes arguments of a starting article title and some integer, $\mu$, and can be described as follows:
\begin{enumerate}
	\item Given an initially empty set $V$, add the root Wikipedia article title $t$.
	\item If $\mu>0$, then repeat the process for each Wikipedia article link $l$ in the article provided by $t$ but for $\mu-1$ and with {the now} non-empty $V$.
	\item Upon collecting all Wikipedia {article titles} in the scraping process, then pull the text data from each Wikipedia article in $V$ and output the entirety of this data to a text file to be used for training.
\end{enumerate}
    	
\change{{\bf Chosen Conceptual Nodes.}} For our work, we consider the following set of starting article titles on Wikipedia and perform the described procedure for each starting article: ``celebrity", ``identity (social science)", ``news", ``organization", ``politics", and ``religion". These Wikipedia starting articles correspond with a single conceptual node and are used to generate corresponding text files to train corresponding fastText models to classify targeted interests into conceptual nodes. It is intuitively obvious that the run-time of this algorithm exponentially increases with respect to $\mu$. We can grossly consider this algorithm to have a run-time of approximately $O(n^\mu)$ where $n$ is the largest number of linked articles among all the articles explored.
    
\change{{\bf Tuning \bm{$\mu$} Parameter.}} It is important to note that this algorithm may require trial-and-error. While scraping, we found the training text data generated when $\mu=0$ to result in models that produced seemingly random conceptual node classifications. Intuitively, if we increase $\mu$ to $\mu=1$, the fastText models would be provided with more text and thus more accurate results would be produced. We found that in the case of $\mu=1$, 92.857\% of targeted interests were classified as belonging to the same conceptual node. However, upon letting $\mu=2$, we saw more appropriate results and a more expected distribution of conceptual node assignments. \underline{Note}: scraping for when $\mu=2$ took several days to finish. For context, the machine we performed scraping {with} was equipped with an Intel Core i7-7700 quad-core processor with 32 GB of RAM.

\change{
\subsection{Application}
A common challenge of machine learning is that trained models are often impossible to interpret. In other words, when you have a machine learning model (e.g., an artificial neural network) undergo training, it is difficult --- if not impossible --- to get a useful understanding of the derived model. Works investigating issues pertaining to the interpretation issue of ML models are known under the field of {\it explainable artificial intelligence} (XAI)~\cite{xai1, xai2}. An advantage from our approach considering conceptual nodes is that it provides a framework for advertisement designers to intuitively select their own set of parameters that make sense for their marketing aims. From there, these conceptual nodes can leverage machine learning models to learn relationships between them and how they impact user click responses. This allows marketers make helpful insights about what themes resonate with users. This kind of insight may be more difficult to attain using a ``black box" machine learning model. An obvious alternative would be to consider approaches for topic modelling --- via some topic model such as Pachinko Allocation Model (PAM)~\cite{li2006pachinko, mimno2007mixtures}. However, the main function of topic models is to extract abstract thematic similarities from text. The goal of conceptual nodes are to be used as input features to predict CTR for a given advertisement. An obvious next step for this work would be to incorporate topic models for selecting conceptual nodes. However, we are most interested in demonstrating the viability of this approach to CTR prediction to show that reasonable CTR prediction can be accomplished with no user-specific data. Also of note is that our approach to CTR prediction allows for entities outside an OSN platform to analyze their own advertisements on these platforms for CTR analytics without relying on predictions performed by the platform itself. 
}

\section{Learning Models} \label{sec:learning-models}
A prominent aim of this work is to explore methods that effectively approximate the CTR of advertisements on Facebook based on the conceptual nodes of targeted interests. To do this, we employ the following learning models:  AdaBoost regressor (ABR)~\cite{adaboost1}, 
decision tree regressor (DTR)~\cite{dtr1}, multi-layer perceptron regressor  (MLP)~\cite{mlp1}, 
and random forest regressor (RFR)~\cite{rfr1}. 
For this work, we use the implementations of these machine learning models supplied by the SciKit-Learn API~\cite{scikit-learn}. 
The ABR model fits a regressor on an original data-set, fitting additional copies under the same regressor, while adjusting weights of instances according to error of the current prediction. The DTR model fits data under a sine curve and learns local linear regressions by approximating the sine curve. The MLP model is a supervised model that learns a non-linear function $f(\cdot):R^m \rightarrow R^o$ by training on a data-set under a set of provided features. The RFR model is an ensemble technique that incorporates the notion of ``bagging" across training examples and features when training a set of decision trees, rather than a single decision tree, to perform regression.
	
We also consider three cases of input parameters to fit the models against the real-world number of clicks for each advertisement in the set of advertisements. That said, we consider the three cases of input:
For the input features for our machine learning models, we consider three cases of input:
\begin{enumerate}[label=(\Alph*)]
	\item {\it counts of each conceptual node}
	\item {\it spend (Russian rubles) \& counts of each conceptual node}
	\item {\it spend, days online, \& counts of each conceptual node}
\end{enumerate}
The decision to consider three cases of input for our machine learning models was motivated by the initiative to see how our conceptual nodes, on their own, compare to including other {simple} features for our learning models in terms of CTR prediction accuracy. We discuss the accuracy of the learning models, under each case, in Section~\ref{sec:results}.
In each of the three cases considered, we consider the input provided to the learning models to be vectors. For instance, in Case A we consider the input vector $[0, 1, 0, 2, 0, 0]$ to be the combination of the conceptual nodes --- corresponding respectively with {\it celebrity}, {\it identity}, {\it news}, {\it organization}, {\it politics}, and {\it religion}. For Case B, an additional leftmost element is included for {\it spend}; for Case C, two additional leftmost elements are included for {\it spend} and {\it days online} respectively. 

\begin{figure*}
    \centering
    \subfloat[]{\includegraphics[width=0.33\linewidth]{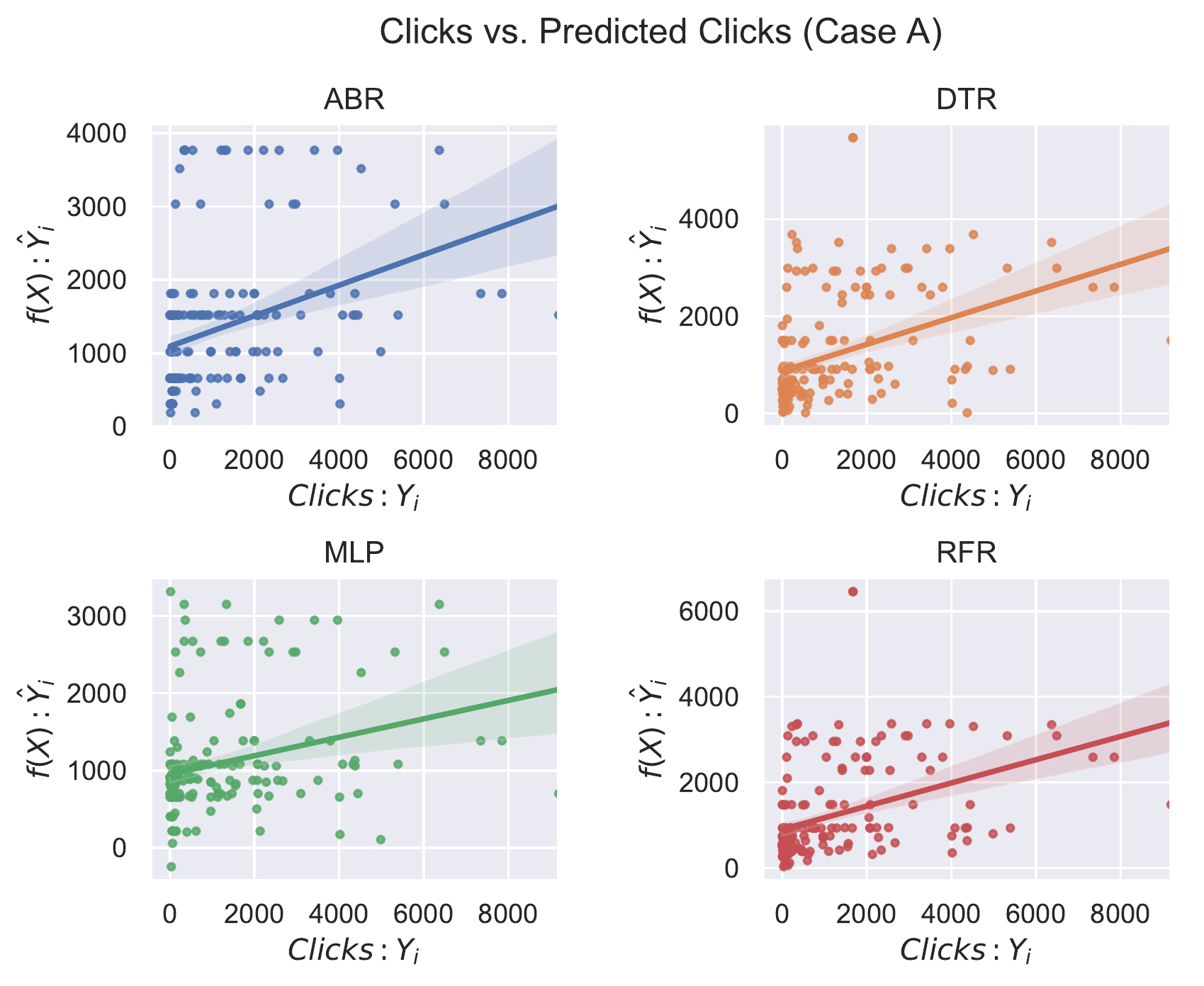}}
    \subfloat[]{\includegraphics[width=0.33\linewidth]{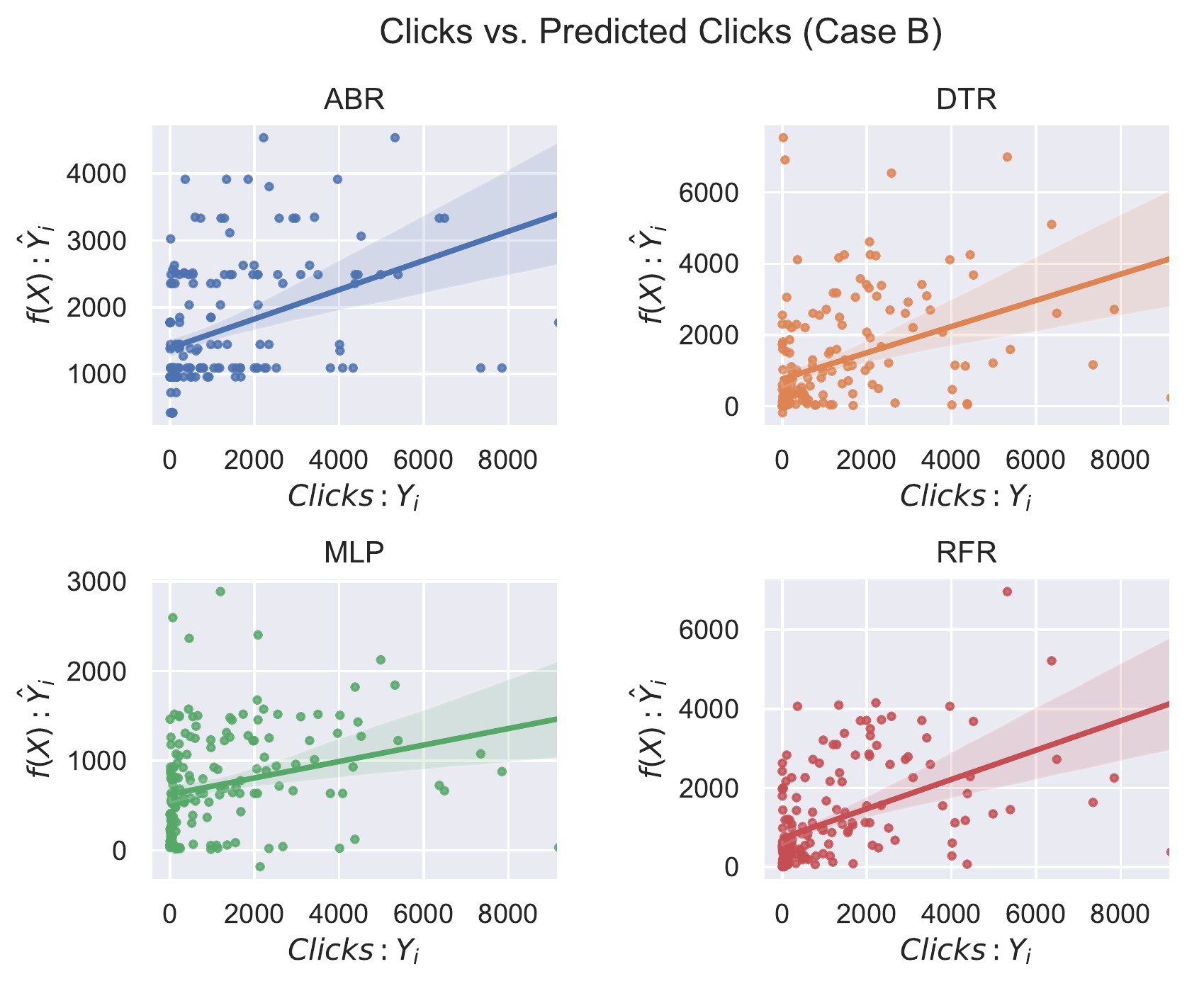}}
    \subfloat[]{\includegraphics[width=0.33\linewidth]{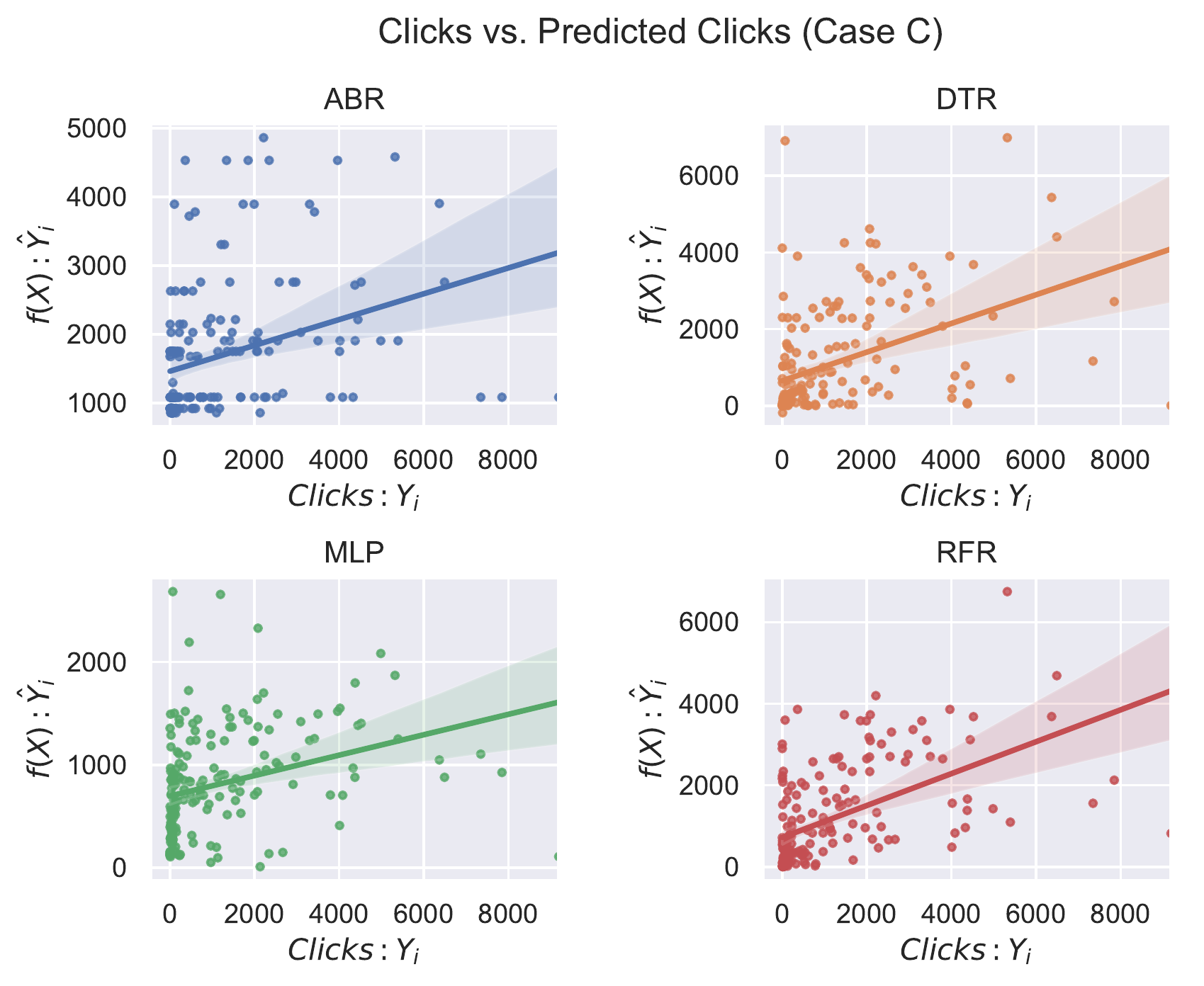}}
    \caption{Scatter plots of real (x-axis) vs. predicted (y-axis) clicks for each model under each input case.}
	\label{fig:models-accuracy}
\end{figure*}

\section{Optimal Conceptual Node {Combinations}} \label{sec:node-allocation}
{Here, we present the formulation for finding optimal combination of conceptual nodes that maximizes the CTR under a given machine learning model trained to perform CTR prediction. }
We use a brute force algorithm to observe the optimal allocation of conceptual nodes given a budget $b$, where $b$ is the number of conceptual nodes to allocate. However, the algorithmic complexity of a brute force approach is too expensive to apply in real-world settings for large values of $b$. The algorithmic complexity of a brute force algorithm for this problem is $\Theta(b^{|\mathcal{N}|})$, where $\mathcal{N}$ is the set of conceptual nodes. To alleviate this shortcoming, w
{We then present a greedy algorithm and a genetic algorithm that run in linear time.} 
%
%
It is worth mentioning that the proposed algorithms for conceptual node allocation only consider learning models under Case A. 


{\bf Optimization Formulation.}
The Optimal Conceptual Node Combination (OCNC) problem can be formally described using an {\it integer nonlinear program} (INLP) formulation. The INLP formulation, provided below, considers a value function $f(\cdot)$, a conceptual node combination $C=[C_1, \cdots, C_{\mathcal{|N|}}]$, and a budget~$k$,
    \begin{align}
        \text{maximize }   & f(C) \label{eq:obj-func} \\
        \text{subject to } & C_i \in \mathbb{N} & 1 \leq i \leq \mathcal{|N|} \label{eq:constraint1} \\
                           & \sum_{i=1}^{\mathcal{|N|}} C_i \leq k & k\in\mathbb{N} \label{eq:constraint2}
    \end{align}
where, as a reminder, $\mathcal{N}$ is the set of conceptual nodes and $\mathbb{N}$ is the set of natural numbers (including 0). The objective function $f(C)$ in Eq.~\eqref{eq:obj-func} represents the click prediction function returned by some trained machine learning model. Constraint~\eqref{eq:constraint1} restricts a conceptual node combination to be comprised of numbers belonging to the natural number set (i.e., no negative amount of conceptual nodes). Finally, constraint~\eqref{eq:constraint2} ensures that the number of conceptual nodes allocated to $C$ do not exceed the considered budget $k$. From here, we discuss the hardness of solving this problem under some assumptions regarding the value function $f(\cdot)$.


{\bf NP-Hardness.} The hardness of the OCNC problem is dependent on the class $f(\cdot)$ belongs to --- which relies on which ML model is used, the training data, the training hyperparameters, etc. For example, with a linear regression model, the function is linear and makes the problem trivially easy to solve: simply select whichever conceptual node has the largest coefficient and spend all your budget $k$ on that conceptual node. However, for more advanced models, such linearity is likely not possible. For instance, in MLP models, that can result in a composition of convex/concave functions that can make the resulting function $f(\cdot)$ demonstrate non-convexity/concavity~\cite{nonconvex-ml}. No guarantees can be made about the nature of the resulting $f(\cdot)$. If it demonstrates non-convexity/concavity or some other rigorous behavior, the OCNC problem could then be considered as NP-hard. This can be suggested under the observation that global optimization of a non-convex/concave function is found to be NP-hard~\cite{nonconvex1, nonconvex2}. A robust proof for the models considered in this work is beyond the scope of this paper.

\subsection{Proposed Greedy Algorithm}
The greedy algorithm uses a machine learning model as a heuristic to incrementally allot conceptual nodes. Essentially, it starts off by selecting {0} conceptual nodes, then will generate {children combinations} where {1} more conceptual node is considered (e.g., {$[0, 0, 1]\rightarrow \{[1, 0, 1], [0, 1, 1], [0, 0, 2]\}$}). For each of these {children combinations}, it will predict the number of clicks using the provided model. It then repeats this process for the {combination} that produced the highest predicted click values until it exhausts its budget. The pseudocode for this algorithm can be found in Algorithm~\ref{alg:greedy}. 

\begin{algorithm}[h]
    \small
    \SetKwData{Left}{left} \SetKwData{This}{This} \SetKwData{Up}{up}
    \SetKwProg{Proc}{Procedure}{}{end}
    \SetKwInOut{Input}{In} \SetKwInOut{Output}{Out}
    
    \Input{value function $f$, budget $k$, conceptual node set $\mathcal{N}$, current combination $C$ (initially empty).}
    \Output{Solution to OCNC problem.}

    \BlankLine
    
    \Proc{\textsc{Children}$(nodes)$}{
        $C \gets \{\}$\;
        \For{$i \gets 1$ to $|nodes|$}{
            $c \gets \text{copy of }nodes$\;
            $c_i \gets c_i +1$\;
            $C \gets C \cup c$\;
        }
        \Return $C$\;
    }

    \BlankLine

        \If{$k \leq 0$}{
            \Return $C$\; 
        }
        \If{$C = \emptyset$}{
            $C \gets [0,0,\cdots,0]$ such that $|C|=\mathcal{|N|}$\;
        }
        \tcc{\text{\textsc{Children}}$([0, 0, 1])\rightarrow \{[1, 0, 1], [0, 1, 1], [0, 0, 2]\}$}
        $C' \gets \textsc{Children}(C)$\;
        $child \gets \argmax_{c \in \text{C'}}f(c)$\;
        \Return $\textsc{Greedy}(f, k-1, \mathcal{N}, child)$\;
    
	\caption{\small Greedy algorithm.}
	\label{alg:greedy}
\end{algorithm}

\change{
{\bf Complexity of the proposed Greedy Algorithm.} 
The complexity of Algorithm~\ref{alg:greedy} depends closely on the complexity of the value function $f$. Assuming that $f$ exhibits a constant runtime complexity $O(1)$, then the resulting runtime complexity for Algorithm~\ref{alg:greedy} is $O(k \cdot \mathcal{|N|})$. If the value function $f$ exhibits any other runtime complexity, then the runtime of Algorithm~\ref{alg:greedy} becomes $O(k\cdot \mathcal{|N|} \cdot O(f))$.
}

\subsection{Genetic Algorithm}
We implemented a genetic algorithm for conceptual node {allocation} that is given a trained learning model for its fitness function. Additionally, the genetic algorithm is given a budget $k$ for node allocation to ensure that mutated DNA strands and randomly generated children in an initial population do not exceed the {budget for} conceptual nodes considered. 
For our genetic algorithm, we considered $G=1000$ generations and populations of $P=35$ individuals. 
No advanced hyper-parametric tuning was performed to arrive at these parameters. 
We do not provide the genetic algorithm in this paper because the {advent of genetic algorithms is} general enough to be applied to a wide variety of {problems~\cite{genetic}.}

\change{{\bf Complexity of the proposed Genetic Algorithm.} Generally, the runtime complexity of genetic algorithms are dependent on the number of generations, making the runtime $O(P \cdot G \cdot O(f))$ where $P$ is the population size, $G$ is the number of generations, and $O(f)$ is the runtime of the trained model. Each initial/mutated individual $c$ in a population of size $\mathcal{|N|}$ in a population at any point in time must meet the condition $\sum_i^{\mathcal{|N|}} c_i \leq k$. Because we modified the genetic sub-functions (e.g., {\tt mutate}) such that no individuals that violate this constraint are produced, the budget parameter~$k$ does not impact the complexity of the algorithm.}


\change{\subsection{Runtime Comparison.} In practice, we observed the Greedy algorithm is generally faster than the Genetic algorithm. However, we show in Section~\ref{sec:results} that the Genetic algorithm achieves greater predicted click values. The runtime performance of the Genetic algorithm can be reduced, though, by tuning the hyper-parameters $P$ and $G$ --- though this is likely to also reduce the quality of the produced conceptual node combinations.}


\begin{table}
	\caption{Pearson correlation coefficients of each model under each input case considered in this work.}
	\label{table:corr-coeff}
	\centering	
	\begin{tabular}{ccccc}
		\toprule
		\textbf{Input Case} & \textbf{ABR} & \textbf{DTR} & \textbf{MLP} & \textbf{RFR} \\
		\midrule
		A & 0.184 & 0.357 & 0.199 & 0.427 \\
		B & 0.167 & 0.687 & 0.406 & 0.681 \\
		C & 0.508 & 0.724 & 0.052 & 0.755 \\
		\bottomrule
	\end{tabular}
\end{table}

\begin{table}
	\caption{Average allocation for each conceptual node for the top 10\% of advertisements CPR values.}
	\label{table:click-spend}
	\centering
	\begin{tabular}{cc}
		\toprule
		\textbf{Conceptual Node} & \textbf{Average Allocation} \\
		\midrule
		Celebrity    & 1.116809 \\
		Identity     & 0.495726 \\
		News         & 0.054131 \\
		Politics     & 1.712251 \\ 
		Organization & 1.484330 \\
		Religion     & 0.384615 \\
		\bottomrule
	\end{tabular}	
\end{table}

\section{Results} \label{sec:results}
Here we discuss the results of this work. More specifically, we are interested in following points: (A) the accuracy of the machine learning models, (B) the performance of the proposed algorithms for conceptual node allocation, and (C) insights on the conceptual nodes that are most influential in attracting clicks from users.

\subsection{Accuracy of Machine Learning Models}
In order to make the case that our approach to CTR prediction using our conceptual nodes has merit, we must affirm that at least one of the trained learning models used to predict clicks is reasonably close to real click values. To verify this, we train each of the four considered learning models on a random set consisting of 95\% of the IRA data, dedicating the remaining 5\% of data to testing accuracy.
Figure~\ref{fig:models-accuracy} shows the accuracy of the four learning models, under Case C, in terms of how predicted click values $\hat{Y}_i$ correlate with the real click values $Y_i$ for each advertisement $i$ in the {testing subset of advertisements}. In these plots, it can be seen that each model maintains a positive correlation between predicted and real click values. To more thoroughly analyze the accuracy of these models, Table~\ref{table:corr-coeff} provides the Pearson correlation coefficients between $Y_i$ and $\hat{Y}_i$ for each model under each case. We see significant gains in terms of correlation under Case C for all learning models {with the exception of MLP}. The RFR and DTR models feature the highest correlation coefficients, demonstrating their reliability for CTR prediction in this approach. {Further, across all three cases, DTR and RFR demonstrate the highest efficacy in terms of Pearson correlation coefficient.}

\begin{figure}
	\centering
	\includegraphics[width=\linewidth]{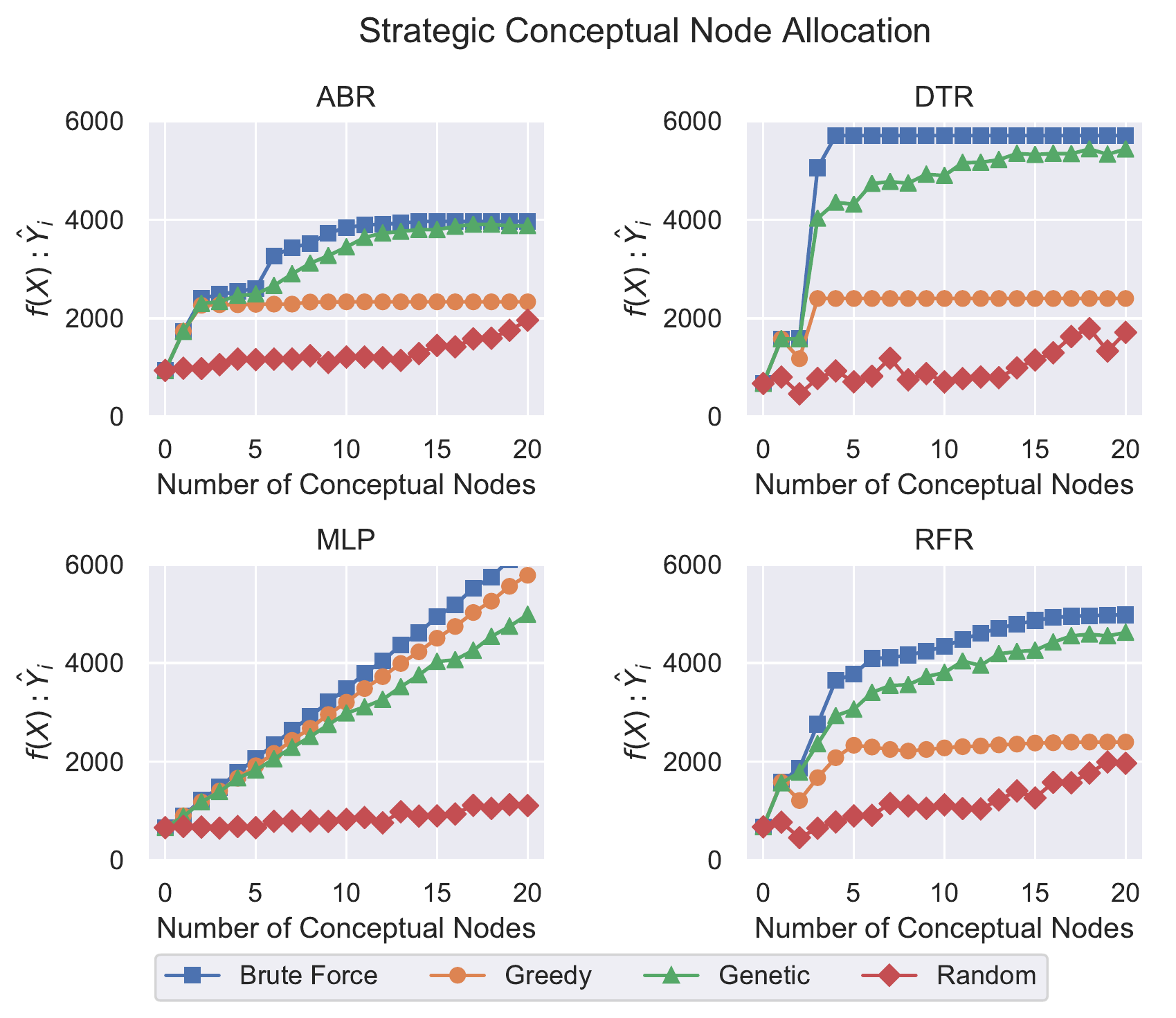}
	\caption{These plots demonstrate the effectiveness of the strategic conceptual node allocation algorithms considered for this work under Case A.}
	\label{fig:models_performance}
\end{figure}

\subsection{Performance of Proposed Algorithms}
{To reiterate, this approach to CTR prediction is unorthodox when compared to other approaches in the literature. Standard approaches to CTR prediction typically incorporate heavy use of user-based input features. Since rich user input features are difficult to get for entities other than OSN platforms themselves, this is an inaccessible approach for most researchers. Having said that, to the best of our knowledge, there are not any reasonable baseline algorithms to compare our approach against. As such, we} compare our greedy and genetic algorithms against two other algorithms, brute force and random allocation. Brute force allocation will serve as our upper bound. It simply runs through all possible allocations of conceptual nodes and returns the allocation that provides the highest click prediction. To reiterate, the run-time for this algorithm is $\Theta(k^{|\mathcal{N}|})$ {where $\mathcal{N}$ is the set of conceptual nodes} --- making it very computationally expensive. Our random algorithm simply {allocates} a random allocation of conceptual nodes within budget, making it the lower bound for comparison.
	
First, we train our learning models with {\it all} of the IRA data. For each budget value $k=0,1,\cdots,20$, we allocate conceptual {nodes for $k$} using each of the four algorithms and then predict CTR using each learning model. We keep track of the CTR predictions for each value of $k$ under each model to generate a curve that will be used to plot the performance of the algorithms in maximizing clicks. To adequately compare the results of this process, we perform this task 100 times and average the results over the 100 iterations to account for any random behaviors when training our learning models.
	
Under {the ABR, DTR, and RFR learning models}, we observe that the genetic algorithm performs the best among the three non-brute-force solutions --- with performance being very close to the optimal brute force {solution}. Under the MLP model, we observe that the greedy algorithm outperforms the genetic algorithm and performs {{\it very}} close to the optimal brute-force solution. 
The performance of these algorithms under the four learning models and Case A, can be seen in Figure~\ref{fig:models_performance}. {It is worth noting that the genetic algorithm reaches near-optimal solutions with the learning models that exhibit the highest accuracy in terms of Pearson correlation coefficients.}

\subsection{Most Influential Conceptual Nodes}
For this work, we wanted to be able to draw some
{social} intuition as to {\it which} conceptual nodes among the six considered proved to be more influential in attracting clicks from users. { First, we normalize the data to account for spend. This is necessary because advertisements with more spend are prioritized more in Facebook's system.} %
	%
To {do this, we disregard all advertisements that had 0 rubles spent on them. From there}, we normalize values for spend and clicks {of the remaining advertisements} by 
{grabbing the top 10\% of advertisements with the most clicks per spend.}

Table~\ref{table:click-spend} provides the {average occurrence of each conceptual node among the top 10\% of clicks-per-ruble advertisements}.
From these results, we can clearly observe that the conceptual nodes of {\it politics}, {\it organization}, and {\it celebrity} are the most influential conceptual nodes by a considerable margin. Interestingly, the {\it news} conceptual node has very insignificant influence according to these results.

\section{Conclusions \& Future Directions}
\label{sec:conclusion}
In this work, we perform an content-aware approach to CTR prediction using machine learning models with our notion of conceptual nodes and a relatively small data-set. Through this approach for CTR prediction, our learning models achieve impressive efficacy. Also, we introduce a greedy algorithm and implement a genetic algorithm to find optimal conceptual node combinations to solve the considered optimization problem under our trained learning models, with our genetic algorithm performing near optimal solutions under three of our four models. Lastly, we acquire some {social} insights that {\it politics}, {\it celebrity}, and {\it organization} are the most influential conceptual nodes for attracting user clicks while {\it news} is the least effective.
\change{This work considered a static set of conceptual nodes to perform click maximization for smart advertisement. An obvious next step for this work would be to incorporate state-of-the-art topic models to dynamically generate an initial set of conceptual nodes. However, for the context of this work, that was not a key emphasis. 
}

\bibliographystyle{ieeetr}
\bibliography{refs}

\end{document}